\documentclass{MYws-procs10x7}
\def\bc{\begin{center}}
\def\ec{\end{center}}
\def\beq{\begin{equation}}
\def\eeq{\end{equation}}
\newcommand{\bmath}{\begin{displaymath}}
\newcommand{\emath}{\end{displaymath}}
\newcommand{\beqn}{\begin{eqnarray}}
\newcommand{\eeqn}{\end{eqnarray}}
\newcommand{\beqns}{\begin{eqnarray*}}
\newcommand{\eeqns}{\end{eqnarray*}}
\newcommand{\ba}{\begin{array}{c}}
\newcommand{\bat}{\begin{array}{cc}}
\newcommand{\ea}{\end{array}}

\newcommand{\nn}{\nonumber}

\newcommand{\pr}{Phys.~Rev. }
\newcommand{\np}{Nucl.~Phys. }
\newcommand{\pl}{Phys.~Lett. }
\newcommand{\prl}{Phys.~Rev.~Lett. }

\newcommand{\cL}{{\cal L}}

\newcommand{\lsim}{\stackrel{<}{_\sim}}

\def\eqn#1{(\ref{#1})}

\begin{document}

\title{$\mathbf{\varepsilon'/\varepsilon}$ \ in The Standard Model\hskip 1pt:\
Theoretical Update}


\author{A. Pich}

\address{Departament de F\'{\i}sica Te\`orica, IFIC,
Universitat de Val\`encia--CSIC,\\
E-46071 Val\`encia, Spain. \ E-mail: Antonio.Pich@ific.uv.es}

\twocolumn[\maketitle\abstract{
A complete analysis of isospin breaking in $K\to 2\pi$ amplitudes, including
both strong ($m_u\not= m_d$) and electromagnetic corrections at next-to-leading
order in chiral perturbation theory, has been achieved recently.\cite{CENP:03c}
We discuss the implication of these effects, together with
the previously known chiral loop corrections,\cite{PPS:01}
on the direct CP-violating ratio $\varepsilon'/\varepsilon$.
One finds \
$\mathrm{Re}\left(\varepsilon'/\varepsilon\right) =
\left( 19\, {}_{-\, 9}^{+11}\right) \cdot 10^{-4}$.
}]

\section{Introduction}

The CP--violating ratio  $\varepsilon'/\varepsilon$  constitutes
a fundamental test for our understanding of flavour--changing
phenomena.
The experimental status has been clarified by the
KTEV,\cite{KTEV:03}
${\rm Re} \left(\varepsilon'/\varepsilon\right) =
(20.7 \pm 2.8) \cdot 10^{-4}$,
and NA48,\cite{NA48:02}
${\rm Re} \left(\varepsilon'/\varepsilon\right) =
(14.7 \pm 2.2) \cdot 10^{-4}$,
measurements.
The present world average,\cite{KTEV:03,NA48:02,NA31,E731}
\beq\label{eq:exp}
{\rm Re} \left(\varepsilon'/\varepsilon\right) =
(16.7 \pm 1.6) \cdot 10^{-4} \, ,
\eeq
demonstrates the existence of direct CP violation in K decays.

The CP violating signal is generated through the
interference of two different $K^0\to\pi\pi$ decay amplitudes,
\beq\label{eq:eps'}
{\varepsilon^\prime\over\varepsilon} =
\; e^{i\Phi}\; {\omega\over \sqrt{2}\,\vert\varepsilon\vert}\;\left[
{\mbox{Im} A_2\over\mbox{Re} A_2} - {\mbox{Im} A_0\over\mbox{Re} A_0}
 \right] \, .
\eeq
In the limit of CP conservation, the isospin amplitudes $A_{0,2}$ 
are real and positive.
$\varepsilon'/\varepsilon$ is
suppressed by the small ratio
$\omega = \mbox{Re} A_2/\mbox{Re} A_0 \approx 1/22$.
The strong S--wave rescattering of the two final pions generates a
large phase-shift difference between the two amplitudes,
making the phases of $\varepsilon'$ and $\varepsilon$ nearly equal:
$\Phi \approx \delta_2-\delta_0+ \pi/4\approx 0$.
Thus, unitarity corrections\cite{PPS:01} 
play a crucial role in $\varepsilon'/\varepsilon$.
Moreover,
the ratio $1/\omega$ amplifies any potential contribution
to $A_2$ from small isospin-breaking corrections induced by $A_0$.

The CP--conserving amplitudes $\mbox{Re} A_I$, their ratio
$\omega$ and $\varepsilon$ are usually set to their experimentally
determined values. A theoretical calculation is only needed
for $\mbox{Im} A_I$.

\section{Theoretical Framework}
\label{sec:theory}

Owing to the presence of very different mass scales
($M_\pi < M_K \ll M_W$), the gluonic corrections to
the $\Delta S=1$ process
are amplified by large logarithms.
The short-distance logarithmic corrections can be summed up using the
Operator Product Expansion (OPE) and the renormalization
group, all the way down from $M_W$ to scales $\mu < m_c$.
One gets 
an effective Lagrangian, defined in the
three--flavour theory,\cite{GW:79,BURAS}
\beq\label{eq:Leff}
\cL_{\mathrm{eff}}^{\Delta S=1}= - \frac{G_F}{\sqrt{2}}\,
 V_{ud}^{\phantom{*}}V^*_{us}\,  \sum_{i=1}^{10}
 C_i(\mu) \, Q_i (\mu) ,
\eeq
which is a sum of local four--fermion operators $Q_i$,
modulated
by Wilson coefficients $C_i(\mu)$ which are functions of the
heavy masses ($M>\mu$) and CKM parameters.
These coefficients are known at the next-to-leading
logarithmic order.\cite{buras1,ciuc1} This includes all
corrections of $O(\alpha_s^n t^n)$ and
$O(\alpha_s^{n+1} t^n)$, where
$t\equiv\ln{(M_1/M_2)}$ refers to the logarithm of any ratio of
heavy mass scales $M_1,M_2\geq\mu$.

To a very good approximation, only two operators are
numerically relevant for $\varepsilon'/\varepsilon$:
the QCD penguin operator $Q_6$ governs $\mbox{Im}A_0$,
while $\mbox{Im}A_2$
is dominated by the electroweak penguin operator $Q_8$.
A naive vacuum insertion approximation to their
hadronic matrix elements results in a large numerical
cancellation,  
leading\cite{munich,rome} to unphysical low values of
$\varepsilon'/\varepsilon$ around $7\times 10^{-4}$.
The true Standard Model prediction is then very sensitive to the
precise values of these two matrix elements.

Below the resonance region
one can use symmetry considerations to define another
effective field theory in terms of the QCD Goldstone bosons.
Chiral perturbation theory ($\chi$PT) describes\cite{WE:79,GL:85}  
the pseudoscalar--octet dynamics, through a perturbative expansion
in powers of momenta and quark masses
over the chiral symmetry breaking scale
$\Lambda_\chi\sim 1\; {\rm GeV}$.
Chiral symmetry fixes the allowed operators.
At lowest order,   
the most general effective bosonic Lagrangian
with the same $SU(3)_L\otimes SU(3)_R$ transformation properties
as $\cL_{\mathrm{eff}}^{\Delta S=1}$
contains three terms:
%
\beq\label{eq:L2}
{\cal L}_{2}^{\Delta S=1}=\,
G_8 \,\cL_8 + G_{27} \,\cL_{27} + G_{\mathrm{ew}} \,\cL_{\mathrm{ew}}\, .
\eeq
$\cL_{\mathrm{ew}}$   
gives the low--energy realization of $Q_8$,
while $Q_6$ is included in the octet term.

$\cL_2^{\Delta S=1}$ determines the $K\to\pi\pi$ amplitudes at $O(p^2)$.
The calculation of the chiral couplings $G_i$
from the short--distance Lagrangian \eqn{eq:Leff}
requires to perform the matching between the two effective theories.
This can be done in the limit of an infinite number of quark colours,
because the four--quark operators factorize into currents
which have well--known chiral realizations.
This is equivalent to the
standard large--$N_C$ evaluations of $\langle Q_i\rangle$.
Therefore, up to minor variations on some input parameters,
the corresponding $\varepsilon'/\varepsilon$
prediction, obtained at lowest order in both the $1/N_C$ and
$\chi$PT expansions, reproduces the published results of the
Munich\cite{munich} and Rome\cite{rome} groups.

\begin{figure}[tb]
\vskip -3pt
\setlength{\unitlength}{0.6mm} \centering  
\begin{picture}(115,113)
\put(0,0){\makebox(115,113){}}
\thicklines
\put(0,101){\makebox(20,13){\large Scale}}
\put(26,101){\makebox(36,13){\large Fields}}
\put(72,101){\makebox(40,13){\large Eff. Theory}}
\put(0,103){\line(1,0){115}} {\large
\put(0,70){\makebox(20,27){$M_W$}}
\put(26,70){\framebox(36,27){$\ba W, Z, \gamma, g \\
     \tau, \mu, e, \nu_i \\ t, b, c, s, d, u \ea $}}
\put(72,70){\makebox(40,27){\vbox{Standard \\ Model}}}
\put(0,35){\makebox(20,18){$\lsim m_c$}}
\put(26,35){\framebox(36,18){$\ba  \gamma, g  \, ;\, \mu ,  e, \nu_i
             \\ s, d, u \ea $}}
\put(72,35){\makebox(40,18){$\cL_{\mathrm{QCD}}^{n_f=3}$,
             $\cL_{\mathrm{eff}}^{\Delta S=1,2}$}}
\put(0,0){\makebox(20,18){$M_K$}}
\put(26,0){\framebox(36,18){$\ba\gamma \; ;\; \mu , e, \nu_i  \\
            \pi, K,\eta  \ea $}}
\put(72,0){\makebox(40,18){$\chi$PT}}
\linethickness{0.3mm}
\put(44,32){\vector(0,-1){11}}
\put(44,67){\vector(0,-1){11}}
\put(48,59.5){OPE}
\put(48,24.5){$N_C\to\infty$}}
\end{picture}
\vskip 5pt
\caption{Evolution from $M_W$ to $M_K$.\protect\cite{EFT}}
\label{fig:eff_th}
\end{figure}
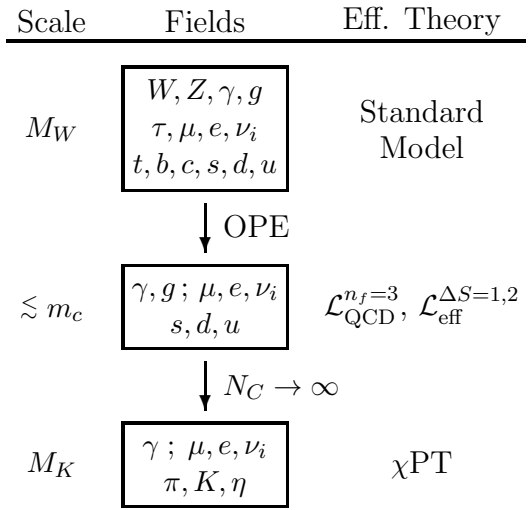


\section{Chiral Corrections}
\label{sec:loops}

The large--$N_C$ limit is only applied to the matching between
the 3--flavour quark theory and $\chi$PT.
The evolution from the electroweak
scale down to $\mu < m_c$ has to be done without any unnecessary expansion
in powers of $1/N_C$; otherwise, one would miss large corrections
of the form ${1\over N_C} \ln{(M/m)}$, with $M\gg m$ two widely
separated scales.\cite{BBG87}

Similarly, the long--distance rescattering of the two pions
generates large logarithmic corrections, through chiral loops
which are of higher order in both the momentum
and $1/N_C$ expansions.\cite{PPS:01}
These next-to-leading contributions, which give rise to
the large S--wave strong phases $\delta_I$
($\delta_I=0$ at $N_C\to\infty$), were overlooked in the   
first large--$N_C$ predictions\cite{munich,rome}
giving a too small $\varepsilon'/\varepsilon$.
Larger values of
$\varepsilon'/\varepsilon$ were in fact obtained within
models containing some kind of pion rescattering.\cite{Trieste}
The $\chi$PT framework allows us to incorporate
rigorously these corrections in a model independent way.

The one--loop $\chi$PT analyses\cite{CENP:03c,PPS:01,KA91}
of $K\to 2 \pi$ show indeed that pion loops provide an
important enhancement of the $A_0$ amplitude, associated with
large infrared logarithms involving the light pion mass,
and a sizeable reduction of $A_2$.
These chiral corrections destroy the numerical
cancellation between the $Q_6$ and $Q_8$ contributions,
generating a large enhancement
of the $\varepsilon'/\varepsilon$ prediction.\cite{PPS:01}

A complete one--loop calculation,
including electromagnetic and isospin violation corrections, has been
achieved recently.\cite{CENP:03c}
The loop contributions are fully determined by chiral symmetry.
The local corrections generated by the different
higher-order chiral lagrangians have been computed at leading
order in the $1/N_C$ expansion.

To account for isospin breaking, we can write
%
\beqn\label{eq:cpiso}
{\varepsilon^\prime\over\varepsilon} = -
{e^{i\Phi}\omega_+\over \sqrt{2}\,\vert\varepsilon\vert}\left[
\displaystyle\frac{\mbox{Im} A_{0}^{(0)} }{\mbox{Re} A_{0}^{(0)} }
(1 - \Omega_{\rm eff}) - \displaystyle\frac{\mbox{Im} A_{2}^{\rm emp}}{\mbox{Re}
  A_{2}^{(0)} } \right]
\nn\eeqn
where the superscript $(0)$ denotes the isospin limit,
$A_{2}^{\rm emp}$ is the electromagnetic penguin contribution to $A_2$,
and\cite{cdg00}
$\omega_+ = \mbox{Re} A_{2}^{+}/\mbox{Re} A_{0}$
(with $\mbox{Re} A_{2}^{+}$ measured in $K^+\to\pi^+\pi^0$)
differs from $\omega = \mbox{Re} A_{2}/\mbox{Re} A_{0}
=\omega_+ \, \left( 1 +  f_{5/2} \right)$
by a pure $\Delta I=5/2$ effect.

The quantity\cite{CENP:03c}
$\Omega_{\rm eff} = \Omega_{\rm IB} - \Delta_0 - f_{5/2}$
contains all isospin breaking effects to leading order.
$\Delta_0$ accounts for isospin breaking corrections
to $A_0$, while the more traditional parameter
$\Omega_{\rm IB}$ parameterizes the
contributions to $\mbox{Im} A_2$ from other four-quark operators
not included in $A_{2}^{\rm emp}$.
%
Taking $\alpha=0$, the isospin breaking is completely dominated by
the $\pi^0$--$\eta$ mixing contribution\cite{EMNP:00}
$\Omega_{\rm IB}^{\pi^0\eta} = 0.16\pm 0.03$.
Electromagnetic effects give sizeable contributions to all three terms,
generating a destructive interference and a smaller
final value\cite{CENP:03c}
for the overall measure of isospin violation in $\varepsilon'$:
\beq\label{eq:omEff}
\Omega_{\rm eff}= 0.06 \pm 0.08\, .
\eeq

\section{Discussion}

Chiral loops generate an important enhancement
($\sim 35\% $) of the isoscalar $K\to\pi\pi$ amplitude
and a sizeable reduction of $A_2$. This effect gets amplified
in the prediction of $\varepsilon'/\varepsilon$, because
at lowest order (in both $1/N_C$ and the chiral expansion) there
is an accidental numerical cancellation between the $I=0$ and $I=2$
contributions. Since the chiral corrections destroy this cancellation,
the final result is dominated by the amplitude $A_0$.
The small value recently obtained\cite{CENP:03c} for $\Omega_{\rm eff}$
reinforces the dominance of the gluonic penguin operator $Q_6$.
Taking this into account and updating all other inputs,\cite{PPS:01}
the Standard Model prediction for
$\varepsilon'/\varepsilon$ turns out to be
\beq\label{eq:finalRes}
\mbox{Re}\left(\varepsilon'/\varepsilon\right) \; =\;
\left(19\pm 2\, {}_{-6}^{+9} \pm 6\right) \cdot 10^{-4}\, ,
\label{eq:final_result}
\eeq
in excellent agreement with the experimental measurement (\ref{eq:exp}).
The first error has been estimated by varying the
renormalization scale $\mu$ between $M_\rho$ and $m_c$.
The uncertainty induced by $m_s$, which has been taken in
the range\cite{ms}
$m_s(2\, \rm{GeV})=110\pm 20\, \rm{MeV}$,
is indicated by the second error.

The most critical step is the matching between the short and long--distance
descriptions, which has been done at leading order in $1/N_C$.
Since all next-to-leading ultraviolet and infrared logarithms have been
taken into account, our educated guess for the theoretical
uncertainty associated with subleading contributions is $\sim 30\% $ (third error).

The control of non-logarithmic corrections at the next-to-leading order
in $1/N_C$ remains a challenge for future investigations.
Several dispersive analyses\cite{Cir:03,KPdR:01,NA:01,BGP:01}
and lattice calculations\cite{lattice} of $\langle Q_8\rangle$
already exist (most of them in the chiral limit). Taking the chiral corrections
into account, those results are compatible with the value used in
\eqn{eq:finalRes}. Unfortunately, the penguin matrix element is more
difficult to compute.
Two recent     
estimates in the chiral limit,
using the so-called {\it minimal hadronic approximation}\cite{HPdR:03}
and {\it X-boson approach}\cite{BGP:01},
find large $1/N_C$ corrections to $\langle Q_6\rangle$.
It would be interesting to understand the physics behind those
contributions and to study whether corrections
of similar size are present for physical values of the quark
masses.
Lattice calculations of $\langle Q_6\rangle$ are still not very
reliable\cite{GP:01} and
give contradictory results\cite{lattice,Staggered}  (often with the wrong sign).

More work is needed to reduce the present uncertainty quoted
in \eqn{eq:finalRes}. This is a difficult task, but progress in this direction
should be expected in the next few years.

\section*{Acknowledgments}
I want to thank V. Cirigliano, G. Ecker, H.~Neufeld, E. Pallante and I. Scimemi
for a very rewarding collaboration.
This work has been supported 
by the EU HPRN-CT2002-00311 (EURIDICE),  MCYT 
(FPA-2001-3031) and Generalitat Valenciana
(GRUPOS03/013 and GV04B-594).


\begin{thebibliography}{99}

\bibitem{CENP:03c}
V. Cirigliano, G. Ecker, H. Neufeld and A.~Pich,
Eur. Phys. J. C33 (2004) 369;
%
\prl 91 (2003) 162001.

\bibitem{PPS:01} E. Pallante, A. Pich and I. Scimemi,
  \np B617 (2001) 441;
%
  E. Pallante and A. Pich,  \prl 84 (2000) 2568;
  \np B592 (2000) 294.

\bibitem{KTEV:03} KTeV collab.,
  \pr D67 (2003) 012005;  
  \prl  83 (1999) 22.

\bibitem{NA48:02} NA48 collab.,
  \pl B544 (2002) 97;
  Eur. Phys. J. C22 (2001) 231;
  \pl B465 (1999) 335.
   
\bibitem{NA31} NA31 collab.,
  \pl B206 (1988) 169;
  \pl B317 (1993) 233.

\bibitem{E731} E731 collab., \prl 70 (1993) 1203.



\bibitem{GW:79} F.J. Gilman and M.B. Wise, \pr D20 (1979) 2392;
   D21 (1980) 3150.

\bibitem{BURAS} A.J.~Buras, hep-ph/9806471.

\bibitem{buras1} A.J.~Buras, M.~Jamin and  M.E.~Lautenbacher,
   \np B408 (1993) 209; \pl B389 (1996) 749.

\bibitem{ciuc1} M.~Ciuchini  et al., \pl B301 (1993) 263;
    Z. Phys. C68 (1995) 239.

\bibitem{munich}
 S.~Bosch et al., Nucl. Phys. B565 (2000) 3;   
 A.J. Buras et al., Nucl. Phys. B592 (2001)~55;  
%
A.J. Buras and M.~Jamin, JHEP 01 (2004) 048.
%
%

\bibitem{rome}
  M. Ciuchini et al., hep-ph/9910237.

\bibitem{WE:79} S. Weinberg, Physica 96A (1979) 327.

\bibitem{GL:85} J. Gasser and H. Leutwyler, Nucl. Phys. B250
  (1985) 456; 517; 539.


%
%

\bibitem{BBG87} W.A. Bardeen, A.J. Buras and J.-M. G\'erard,
  Nucl. Phys. B293 (1987) 787; Phys. Lett.  
  B192 (1987) 138, B180 (1986) 133.

\bibitem{EFT} A. Pich, hep-ph/9806303.

\bibitem{Trieste}  S.~Bertolini et al., Rev. Mod. Phys. 72 (2000) 65;
 T. Hambye et al., \np B564 (2000) 391.

\bibitem{KA91}
J. Kambor et al., Nucl. Phys. B346 (1990) 17;
  Phys. Lett. B261 (1991) 496; Phys. Rev. Lett. 68 (1992) 1818;
 J. Bijnens, E. Pallante and J.~Prades, Nucl. Phys. B521 (1998) 305;
E. Pallante, JHEP 01 (1999) 012.

%
%
%
%

\bibitem{cdg00}
V.~Cirigliano, J.F.~Donoghue, and E. Golowich,
Eur. Phys. J. C18 (2000) 83.

\bibitem{EMNP:00} G. Ecker, G. M\"uller, H. Neufeld and A.~Pich,
    Phys. Lett. B477 (2000) 88.

\bibitem{ms}
E. G\'amiz et al., JHEP 01 (2003) 060; hep-ph/0408044;
M. Jamin, J.A. Oller and A. Pich, Eur. Phys. J. C24 (2002) 237;
K. Maltman and J. Kambor, Phys. Rev. D65 (2002) 074013;
S.M. Chen et al., Eur. Phys. J. C22 (2001) 31;
H. Wittig, hep-lat/0210025;
C. Aubin et al., Phys. Rev. D70 (2004) 031504;  
M.~G\"ockeler et al., hep-lat/0409312.


\bibitem{Cir:03}
V. Cirigliano et al., \pl B555 (2003) 71, B522 (2001) 245,
B475 (2000) 351; \pr D65 (2002) 054014;
J.F.~Donoghue and E. Golowich,  \pl B478 (2000) 172.

\bibitem{KPdR:01}
M. Knecht, S.~Peris and E. de Rafael, \pl B508 (2001) 117,
  B457 (1999) 227;
S.~Peris and E. de Rafael, \pl B490 (2000) 213.

\bibitem{NA:01} S. Narison, \np B593 (2001) 3.

\bibitem{BGP:01}
J. Bijnens et al., hep-ph/0309216; JHEP 10 (2001) 009,  
06 (2000) 035, 01 (1999) 023.   

\bibitem{lattice}
J.I. Noaki et al., \pr D68 (2003) 014501;
T. Blum et al., \pr D68 (2003) 114506; 
D. Be\'cirevi\'c et al.,  
     Nucl. Phys. B (Proc. Suppl.) 119 (2003) 619.

\bibitem{HPdR:03}
S. Peris, hep-ph/0310063;
T. Hambye, S.~Peris and E. de Rafael, JHEP 05 (2003) 027.

\bibitem{GP:01} M. Golterman and E. Pallante,
   Phys. Rev. D69 (2004) 074503;  
   JHEP 10 (2001) 037, 
   08 (2001) 023.   

\bibitem{Staggered} T. Bhattacharya et al., hep-lat/0409046;
 D. Pekurovsky and G. Kilcup, Phys. Rev. D64 (2001) 074502.

\end{thebibliography}
\end{document}